## QUANTUM MATERIALS

# Bright dipolar excitons in twisted black phosphorus homostructures

Shenyang Huang[1,2]†, Boyang Yu[2]†, Yixuan Ma[2]†, Chenghao Pan[2], Junwei Ma[2], Yuxuan Zhou[2,3], Yaozhenghang Ma[2,3], Ke Yang[4], Hua Wu[2,3,5,6], Yuchen Lei[2], Qiaoxing Xing[2], Lei Mu[2], Jiasheng Zhang[2], Yanlin Mou[2], Hugen Yan[2]*

Bright dipolar excitons, which contain electrical dipoles and have high oscillator strength, are an ideal platform for studying correlated quantum phenomena. They usually rely on carrier tunneling between two quantum wells or two layers to hybridize with nondipolar excitons to gain oscillator strength. In this work, we uncovered a new type of bright infrared dipolar exciton by stacking 90°-twisted black phosphorus (BP) structures. These excitons, inherent to the reconstructed band structure, exhibit high oscillator strength. Most importantly, they inherit the linear polarization from BP, which allows light polarization to be used to select the dipole direction. Moreover, the dipole moment and resonance energy can be widely tuned by the thickness of the BP. Our results demonstrate a useful platform for exploring tunable correlated dipolar excitons.

Dipolar excitons, which are endowed with permanent electrical dipoles, are bosonic particles that have strong repulsive interactions when aligned. They exhibit pronounced nonlinearity (1, 2) and are highly tunable by electrical fields through the Stark effect (3, 4). However, because the existence of a dipole moment requires the separation of the electron and hole, the coupling of the dipolar exciton to the photon is reduced. This is particularly true for interlayer excitons in type II two-dimensional (2D) transition metal dichalcogenide (TMDC) semiconductor heterostructures (5–7), in which electrons reside in one layer and holes in the other, resulting in diminishing oscillator strength.

In coupled quantum wells (8, 9), or in some of the TMDC homobilayers (1–4, 10) and heterobilayers (11–14), one type of carrier can tunnel to the other well or layer, which enables the hybridization of interlayer excitons with bright intralayer ones. This new type of hybrid dipolar exciton gains oscillator strength and can even show up in the light absorption spectra at room temperature (3, 4, 10). They can couple to cavity photons and form exciton polaritons with strong nonlinearity (1, 2), as demonstrated for the K-point hybrid excitons in natural $MoS_2$ bilayers. However, at the K-point of the Brillouin zone for $MoS_2$ and other TMDC semiconduc-tors, the electronic coupling between layers is almost negligible (14, 15). Therefore, the tunneling effect of carriers to form hybrid dipolar excitons sensitively depends on the energy-level resonance condition (11, 16), largely limiting their potential. Moreover, because the resonance energies of hybrid dipolar excitons in TMDC systems are usually in the vicinity of those for bright intralayer excitons, their optical features tend to be overshadowed. Of course, hybrid dipolar excitons that involve carriers in other locations of the Brillouin zone (such as Q and Γ points) in TMDCs exhibit better hybridization with intralayer ones, but they are typically indirect excitons in momentum space, with minimal oscillator strength even for the intralayer exciton themselves (17). Therefore, it remains desirable to have dipolar excitons with large oscillator strength, preferably with new mechanisms for the coexistence of the electrical dipole and optical brightness.

In this study, we reveal such bright dipolar excitons in the infrared regime in 90°-twisted black phosphorus (BP) (18, 19) homostructures. Unlike the dipolar excitons found in TMDC homo- or heterobilayers, our newly discovered excitons are formed by Γ point holes spatially confined to one of the films and electrons extended across both constituent films with new envelope wave functions and energy levels. These excitons are inherently dipolar and exhibit a considerable oscillator strength with an absorption exceeding 1% even at room temperature. Endowed with an out-of-plane electrical dipole, they possess a linear quantum-confined Stark effect. Notably, they exhibit versatile tunability. The dipole orientation can be selected by the light polarization because of the anisotropic atomic structure of BP, and the dipole moment can be tuned from ~0.22 to ~1.1 electron nanometers (e nm) by varying the combination of constituent BP of different thicknesses with a resonance energy ranging from ~0.4 to

1.6 eV. These bright dipolar excitons, with their versatile tunability, are highly desirable for the development of tunable excitonic correlated states and many-body complexes.

## Optical characterization of dipolar excitons

Few-layer BP was exfoliated from a bulk crystal onto a polydimethylsiloxane (PDMS) substrate and then transferred to a substrate (quartz or Si/SiO₂) in sequence to form twisted BP homostructures. The thickness and crystal orientation were determined by infrared extinction spectra (20, 21) [see methods for details (22)]. Figure 1A illustrates the atomic structure of a 90°-twisted BP homostructure composed of three-layer (3 L) BP (top) and four-layer (4 L) BP (bottom), and Fig. 1B shows typical polarized infrared extinction spectra of a homostructure with this configuration. Two new absorption peaks (labeled as I and I') appear below the optical bandgaps of 3 L and 4 L (labeled as $E_{11}$, which represents the interband transition from the topmost valence band to the lowest conduction band), with peak energies of ~0.5 and ~0.59 eV, respectively. Additional new peaks labeled as II and II' can also be observed above the bandgaps of 3 L and 4 L, which originate from higher-energy interband transitions (Fig. 1D). We ascribe these new peaks to bright dipolar exciton resonances. They exhibit considerable oscillator strength even at room temperature, especially for I and I', which both show an extinction of more than 1%. The intralayer exciton resonances ($E_{11}$) from the pristine 4 L and particularly 3 L also appear in the spectra, which are attributed largely to the adjacent unstacked areas and possibly from the stacked but uncoupled regions (for details, see fig. S1). These residual $E_{11}$ signals can serve as indicators of the sample quality, especially given the air sensitivity of few-layer BP (see fig. S4).

From the absorption peak intensities versus the incident light polarization angle in another 90°-twisted 3+4 L homostructure (Fig. 1C), it is clear that dipolar exciton I (I') shares the same linear polarization dependence as $E_{11}$ of 4 L (3 L), that is, along the armchair (AC) direction. Indeed, the absorption of all dipolar excitons is linearly polarized (see figs. S2 and S3), with the polarization along the AC direction of either the bottom or top film; hence, we denote those two groups as excitons I, II, and so on and excitons I', II', and so on. Such inherited linear polarization is a hallmark of the dipolar exciton in BP, offering a new degree of freedom to preset the orientation of the out-of-plane dipole of the exciton, which will be discussed later in the text.

The appearance of a series of new peaks in the spectra indicates strong coupling between the two BP films in the homostructure. The interlayer coupling at the 90°-twisted interface is expected to be different from that between two layers inside pristine BP. Density functional theory (DFT) calculations have predicted that there is an almost negligible interlayer coupling for the states at the Γ point in the valence band but strong interlayer coupling for those in the conduction band at the twisted interface [23] (see fig. S14). The observed inherited polarization direction (either along the top- or bottom-film AC direction, rather than in between) in our experiment serves as direct evidence of the uncoupled valence states in the homostructure [refer to section 3 of the supplementary materials (SM) for details (22)].

[1]Shanghai Frontiers Science Research Base of Intelligent Optoelectronics and Perception, Institute of Optoelectronics, Fudan University, Shanghai 200433, China. [2]State Key Laboratory of Surface Physics, Key Laboratory of Micro- and Nano-Photonic Structures (Ministry of Education), Shanghai Key Laboratory of Metasurfaces for Light Manipulation, and Department of Physics, Fudan University, Shanghai 200433, China. [3]Laboratory for Computational Physical Sciences (MOE), Fudan University, Shanghai 200433, China. [4]College of Science, University of Shanghai for Science and Technology, Shanghai 200093, China. [5]Shanghai Qi Zhi Institute, Shanghai 200232, China. [6]Shanghai Branch, Hefei National Laboratory, Shanghai 201315, China.
*Corresponding author. Email: hgyan@fudan.edu.cn
†These authors contributed equally to this work.

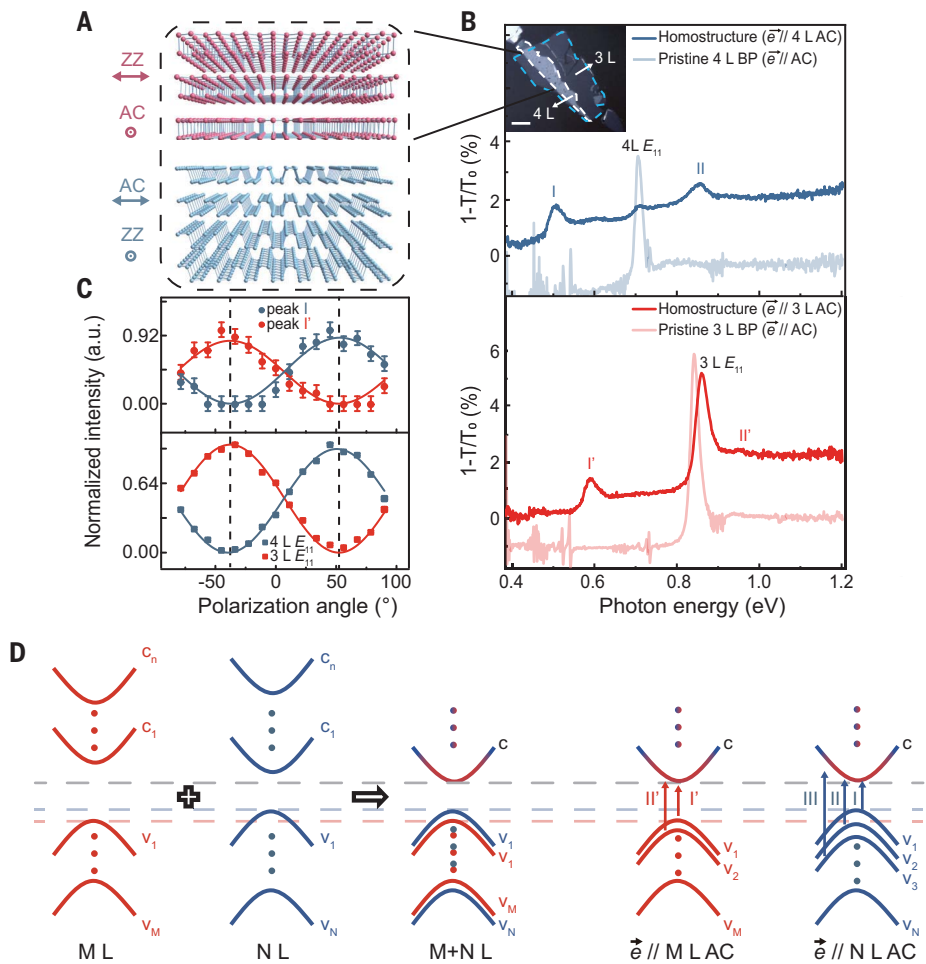

**Fig. 1. Extinction spectra of a 90°-twisted 3+4 L BP homostructure.** (**A**) Illustration of the atomic structure of a 90°-twisted 3+4 L BP homostructure. ZZ, zigzag direction. (**B**) Infrared extinction (1 − T/T₀, where T and T₀ are transmission through the measured sample and bare substrate, respectively) spectra of a 90°-twisted 3+4 L BP homostructure on quartz substrate. The red (blue) curve (vertically shifted for clarity) represents the extinction spectrum of the homostructure under incident light polarized along the 3 L (4 L) AC direction. For comparison, the light red (light blue) curve represents the extinction spectrum of a pristine prestacked 3 L (4 L) BP on PDMS substrate under light polarized along the 3 L (4 L) AC direction. Dipolar excitons are labeled as I, I', II, and II', and the regular exciton associated with the band gap of 3 L (4 L) is labeled as $E_{11}$. The inset is the optical image of this 90°-twisted 3+4 L BP homostructure (scale bar is 10 µm), where the 3 L has a bigger unstacked area, resulting in a strong $E_{11}$ peak in the spectrum. (**C**) The normalized intensities (normalized by the maximal peak height) of optical resonances in pristine prestacked 3 L and 4 L (i.e., $E_{11}$) and the 90°-twisted 3+4 L BP homostructure (i.e., I and I') versus polarization angle. Dots are experimental data, with an error bar of 0.1 for I and I'. Curves are fittings by $I\cos^2(\theta - \alpha)$, where θ is the polarization angle and I and α are fitting parameters. The twist angle of this sample was determined to be 89° ± 2°. a.u., arbitrary units. (**D**) Illustration of the band structure near the Γ point of a 90°-twisted M+N L BP homostructure, together with the typical optical transitions with light polarized along the N L AC or M L AC direction. C, conduction band; V, valence band.

The inequivalent couplings for valence and conduction states at the twisted interface have interesting consequences. First, a series of new conduction bands are formed with energies that are different from those of either of the constituent films, but on the contrary, the valence bands remain intact and are simple combinations of those from the two constituent films (see Fig. 1D). Second, such couplings give rise to spatially extended electrons across the whole homostructure and confined holes either at the top or bottom film (see the bottom panel of Fig. 2A), which form excitons with permanent out-of-plane dipoles and, at the same time, considerable oscillator strengths because electron and hole wave functions still overlap at one of the constituent films. This behavior is in sharp contrast to pure interlayer excitons in most of the TMDC heterostructures and enables us to directly observe the excitons in the room-temperature absorption spectra rather than only in the photoluminescence spectra under low temperature. Third, during the light-matter interaction process, the linearly polarized light along the AC direction of the bottom (top) film will exclusively excite electrons from the valence bands of the bottom (top) film, leaving holes only in the bottom (top) film and electrons all over (Fig. 1D). Therefore, the dipole orientation (either upward or downward) of the photogenerated excitons can be leveraged by the light polarization. Finally, all excitons in the homostructure are dipolar in nature, and they dominate the optical extinction spectra; they are not masked by intralayer excitons (nondipolar). It should be noted that these excitons can be realized under more relaxed experimental conditions. Compared with dipolar excitons in TMDC heterostructures, they exhibit better tolerance to twist-angle misalignment, at least up to a ~15° deviation from the 90° stacking [for details, see section 5 of the SM (22)]. Therefore, we have discovered a truly ideal platform for bright dipolar excitons. In the next sections, we verify these basic properties of dipolar excitons in this platform.

## Linear quantum-confined Stark effect

Endowed with an out-of-plane dipole, the dipolar exciton is expected to exhibit a pronounced linear quantum-confined Stark effect, the fingerprint of the dipolar nature. More intriguingly, the dipole orientation, which is controlled by the polarization of the excitation light, can be directly verified by the applied field. Figure 2A (top panel) is an illustration of a typical device. A 90°-twisted BP homostructure was sandwiched by boron nitride (BN) and few-layer graphene, lightly p-doped silicon was used for the top and bottom infrared transparent electrodes, and BN and SiO₂ were used for the dielectric layer [for details, see methods (22)]. The false-color figures in Fig. 2B display the polarized extinction spectra under a varying applied electric field $F_z$ ($F_z = D/\varepsilon_0$, where D represents the displacement field along the out-of-plane direction and $\varepsilon_0$ is the permittivity of vacuum) for a representative device made of a 90°-twisted 3+4 L BP homostructure with 3 L on the top (see fig. S6 for original extinction spectra). As seen in the top panel of Fig. 2B, with light polarization along the 4 L AC direction, the observed exciton I and II peaks shift almost linearly with the field (upward field is defined as positive), with positive slopes.

Because the dipolar exciton energy shift ($\Delta E$) with a moderate electric field can be expressed as $\Delta E \approx -\boldsymbol{p} \cdot \boldsymbol{E}$, where $\boldsymbol{p}$ is the permanent dipole moment and $\boldsymbol{E}$ is the electrical field ($E_z = F_z/\varepsilon_{BP}$), we can infer that both I and II excitons possess downward permanent dipole moments, consistent with the prediction that holes are localized in the bottom 4 L and electrons are all over. Once we change the light polarization to along the AC direction of the top 3 L, as shown in the lower panel of Fig. 2B, the observed I' exciton exhibits the opposite shift as the I and II excitons, suggesting an upward dipole. Therefore, by leveraging the light polarization, we can selectively generate excitons with the desired dipole orientation.

Within application of a moderate electrical field (−0.6 to 0.6 V/nm), dipolar excitons I and I' exhibit peak shifts larger than 0.1 eV, surpassing the $E_{11}$ peak energy shifts in pure 3 L and 4 L BP. The shifts of $E_{11}$ are supposed to show quadratic dependence (24) on the field and remain mostly unchanged in our field range owing to the lack of a permanent out-of-plane dipole. These shifts are also larger than those observed in pristine 7 L BP or a nontwisted stacking 3 +4 L BP structure (fig. S8). Through a simple linear fitting on the exciton energy dependence on the field, we determined dipole moments $\boldsymbol{p}$ for dipolar excitons I and I' to be ∼−0.7 and ∼0.87 $e$ nm, respectively (the minus sign represents a downward dipole direction), corresponding to separations between electron and hole of 1.4 L and 1.7 L (layer thickness L is ∼0.52 nm). Notably, dipolar exciton II, which has the same dipole orientation as exciton I, exhibits a slower shift rate with electrical field, as shown in fig. S12. Similar results can be also found in other devices (see fig. S7). In addition to the peak energy, the exciton oscillator strength also depends on the field, owing to the enhanced or reduced wave function overlap between the electron and hole within an exciton. As shown in Fig. 2, B and D, I and I' excitons show opposite trends as a result of opposite dipole moments.

To render a precise depiction of the electrical field dependence of the dipolar exciton energy and intensity, we used the 1D chain model [see section 7 of the SM for details (22)]. The model offers a numerical reproduction of energy shifts and intensity changes of dipolar exciton and agrees well with experimental data (see figs. 2, C and D), despite an overestimation in the peak shift of the higher-order exciton II (refer to fig. S12). Our present model captures the dominant features exhibited by the dipolar exciton under an electric field, which can also provide a panoramic view of the quantum confined Stark effect across diverse thicknesses of the 90°-twisted BP homostructure.

### Thickness-controlled out-of-plane dipole moment

Because the electrons are being extended throughout the homostructure while holes are localized at either the top or bottom film, the out-of-plane dipole moment can be deduced from a simple geometric scheme (Fig. 3A). The vertical distance between the central plane of the entire homostructure and that of the top (or bottom) film is the vertical separation of the negative and positive charge centers, hence quantifying the dipole moment. For instance, the out-of-plane dipole moment in 1+1 L is ±0.26 $e$ nm ($\pm e \times 0.5$ L), which increases to 1.04 $e$ nm ($e \times 2$ L) and −0.78 $e$ nm ($-e \times 1.5$ L) for the upward and downward dipolar excitons, respectively, when the thickness increases to 3+4 L. In general, for the 90°-twisted M+N L homostructure (where M and N represent layer numbers, with M L film on the top), the upward and downward dipole moments are $e \times N/2$ L and $-e \times M/2$ L, respectively. This indicates that the out-of-plane dipole moments can be tuned over a wide range by changing the thickness combination, as long as the thickness is smaller than the characteristic size (i.e., Bohr radius) of the exciton, which roughly sets the upper limit of the dipole moment. Such tunability is critical to modify the dipole-dipole interaction and influence the diffusion of exciton ensembles (25, 26). This tunability facilitates the realization of even larger dipoles, which are the key to realizing a more substantial Stark effect.

The thickness-dependent out-of-plane dipole moment has also been verified through the Stark effect. For comparison, another device composed of 5+4 L BP has been studied (5 L BP is on the top). As shown in Fig. 3B, the shift of the downward dipolar exciton in 5+4 L (light polarization along the 4 L AC direction) is obviously faster than that in 3+4 L (polarization along the 4 L AC direction), whereas the shift of upward dipolar exciton in 5+4 L (along the 5 L AC direction) closely matches that in 3+4 L (along the 3 L AC direction). This observation is consistent with the geometrical consideration because, as mentioned above, the downward dipole moment of the former is estimated to be ~−1.3 *e* nm, which is larger than that of ~−0.78 *e* nm for the latter, whereas both the upward dipole moments have an identical value of ~1.04 *e* nm. In experiment, the downward and upward dipole moments in this 5+4 L (3+4 L) device are determined with values of ~−1.1 (−0.7) and ~0.82 (0.87) *e* nm, respectively. Nonetheless, a subtle distinction between these two upward excitonic shifts remains, which is potentially attributable to the additional dipole induced by the field (*24*, *27*). Our 1D chain model reproduced this Stark effect, involving both the permanent and the induced dipoles [see section 7.3 of the SM (*22*)], as depicted by the solid curves in Fig. 3B. Figure 3C summarizes out-of-plane dipole moments in 90°-twisted BP homostructures with various thickness combinations.

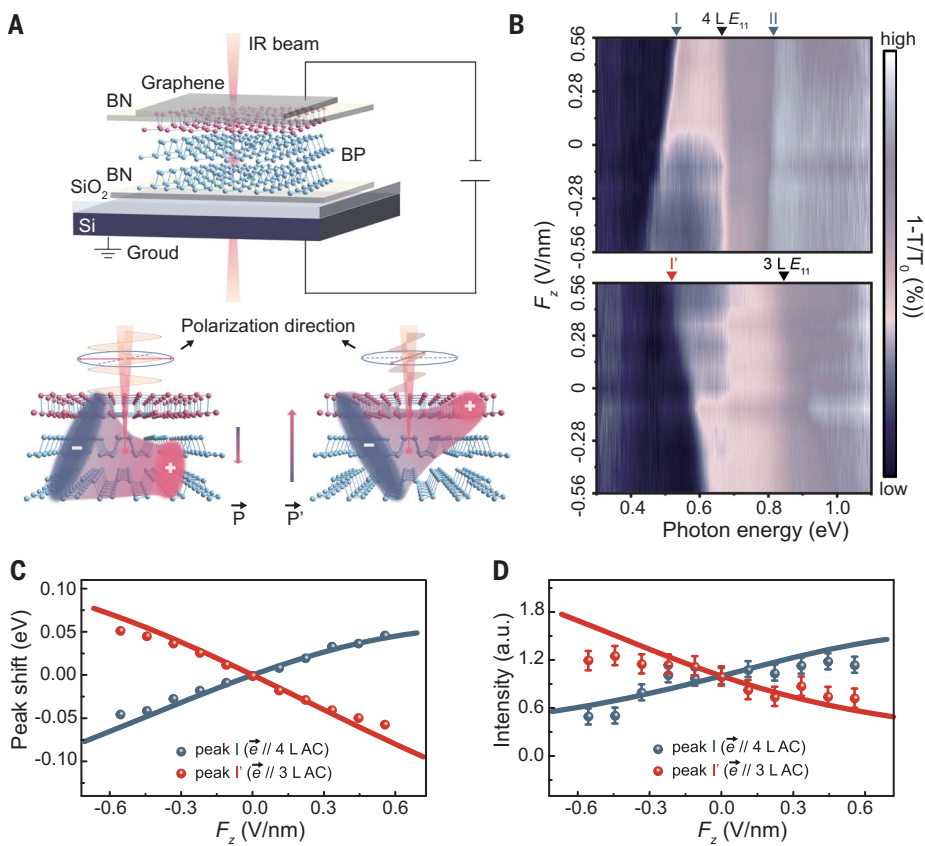

**Fig. 2. 90°-twisted BP homostructure under an electric field.** (**A**) Illustration of a device made of a 90°-twisted BP homostructure for applying a vertical electric field (top) alongside an illustration of the dipolar exciton with the dipole orientation selectable by light polarization (bottom). IR, infrared. (**B**) False-color mapping of extinction spectra with incident light polarized along the 4 L AC (top) and 3 L AC (bottom) directions of a 90°-twisted 3+4 L BP homostructure at different electric fields $F_z$. (**C** and **D**) The peak energy shifts and the normalized intensities (normalized by the intensity without an external electric field) of dipolar excitons I and I' versus electric field, respectively. Dots are experimental data, with an error bar of ∼0.1 for the intensities of I and I' in Fig. 2D. Curves are fitting results based on the model discussed in the SM (22).

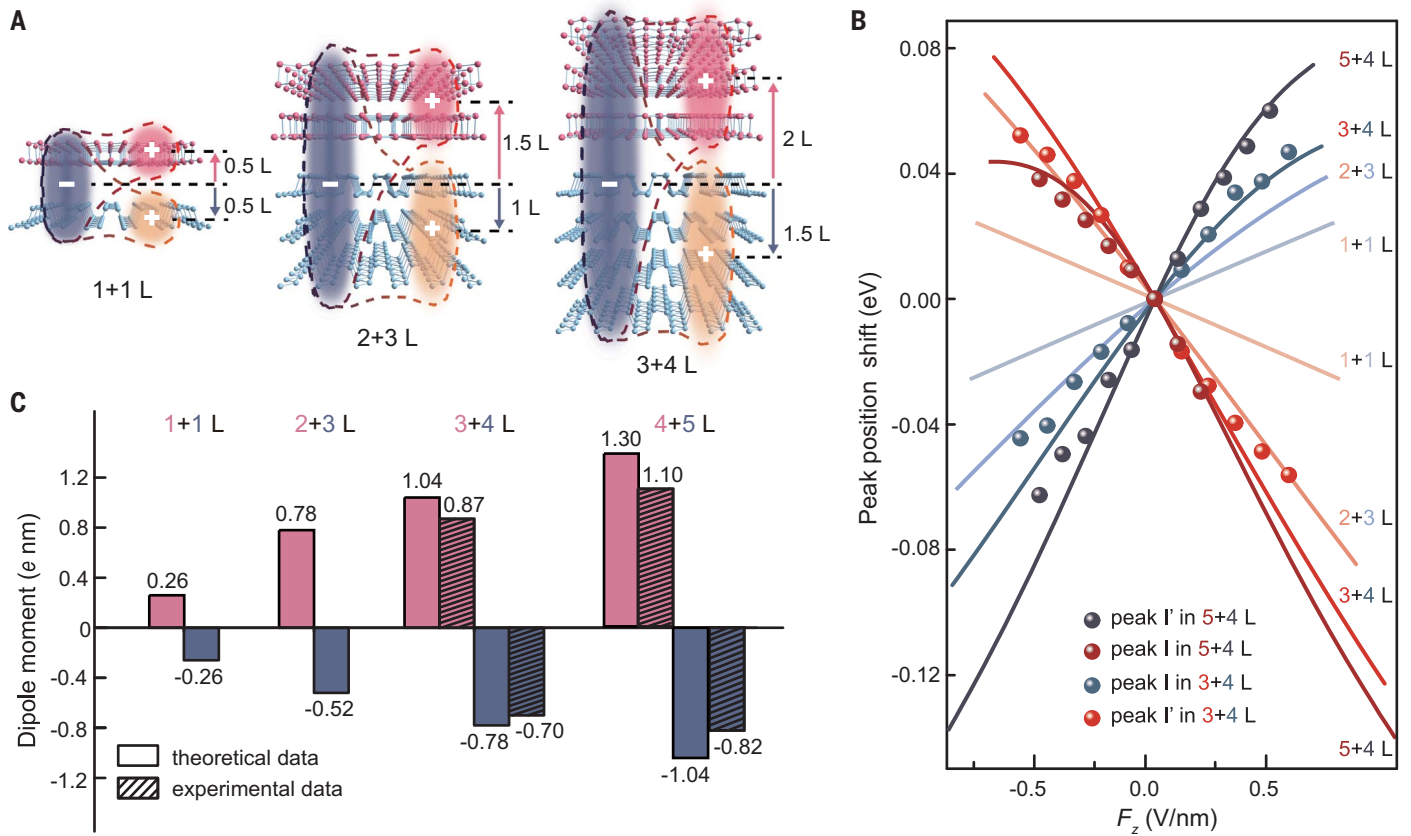

**Fig. 3. Thickness-dependent out-of-plane dipole moment. (A)** Illustration of dipolar excitons with a variety of out-of-plane dipole sizes in 90°-twisted 1+1 L, 2+3 L, and 3+4 L BP homostructures. **(B)** Electric field dependence of the dipolar exciton energies in 90°-twisted BP homostructures with several thickness combinations. Dots are experimental data, and curves are fitting results based on the model discussed in the SM (*22*). The color of the data points and the corresponding fitting curves indicate that the light polarization is along the AC direction of the BP flake, with layer number labeled in the same color. **(C)** The out-of-plane dipole moments in BP homostructures with different thicknesses (the BP flake with pink layer number is shown along the top). The color indicates the polarization along the AC direction of the BP flake, with layer number labeled in the same color. The positive (negative) value represents the upward (downward) dipole orientation.

### Layer-dependent resonance energy of the dipolar exciton

Given the pronounced layer-dependent bandgap at the Γ point in few-layer BP, the dipolar exciton resonance energy is expected to exhibit layer dependence as well. Figure 4, A to C, plots the measured peak energies versus the layer number N for 2+N L, 3+N L, and 4+N L BP homostructures, respectively (see fig. S2 for original extinction spectra). The energies of dipolar excitons in all different branches (e.g., I, I′, II, and so on) redshift with increasing thickness. For instance, dipolar exciton I in 3+N L red shifts from ~0.6 to ~0.4 eV when N increases from three to seven. Our 1D chain model can reproduce this trend well [detailed in section 7 of the SM (*22*)], as shown by dashed curves in Fig. 4. The model also indicates that the dipolar exciton energy can be further extended by choosing different thickness combinations (for example, 1+N L as shown in fig. S9). More specifically, if we only consider the dipolar excitons near the bandgap, they already cover a notable spectral range from ~0.4 to ~1.6 eV, with other high-lying excitons extending the upper spectral limit further. Additionally, the model reveals that the interlayer coupling (γ′c) for the conduction band at the 90°-twisted interface is about 0.60 eV. This value is similar to the γc (~0.55 eV) in pristine few-layer BP. With the obtained value of γ′c, the dipolar ex-citon's energy in a 90°-twisted BP homostructure with any thickness combination can be numerically obtained.

The similar values of γ′c and γc provide an approximation to describe the band structure. We assigned γ′c and γc the same effective value γc^eff, which is to be determined experimentally, and then the dipolar exciton energy for the two groups of excitons in an M+N L homostructure can be directly expressed by

$$E_{n_c n_v}^{M+N}\Big|_{\text{along M (N) L AC}} = E_g - \gamma_c^{\text{eff}}\cos\left(\frac{n_c\pi}{M+N+1}\right) + \gamma_v\cos\left(\frac{n_v\pi}{\overline{M(N)}+1}\right)$$

where $E_g$ is the bandgap of monolayer BP, $\gamma_v$ is the interlayer coupling of the valence band within pristine BP film, $n_c$ ($n_v$) is the index of conduction (valence) band, and $\gamma_c^{\text{eff}}$ is obtained as ~0.54 ± 0.05 eV [for details, see section 7 of the SM (*22*)]. Notably, if we treat the twisted homostructure as a quantum well, the mirror symmetry is broken because of the twisted stacking, in contrast to pristine BP quantum wells. Therefore, interband transition $E_{nm}$ (from the nth valence band to mth conduction band) with $n \neq m$ could occur, which is generally forbidden in pristine few-layer BP. Even though there are possibly several transitions in our spectral range, our experiment reveals that only the $E_{n1}$ interband transition (from nth valence band to lowest conduction band; see Fig. 1D) aligns well with the observed exciton energies. This suggests a dominance of dipolar excitons associated with the $E_{n1}$ interband transitions, and other possible transitions do not show up unambiguously. More compelling evidence favoring the $E_{n1}$ assignment are detailed in section 7.2 of the SM (22).

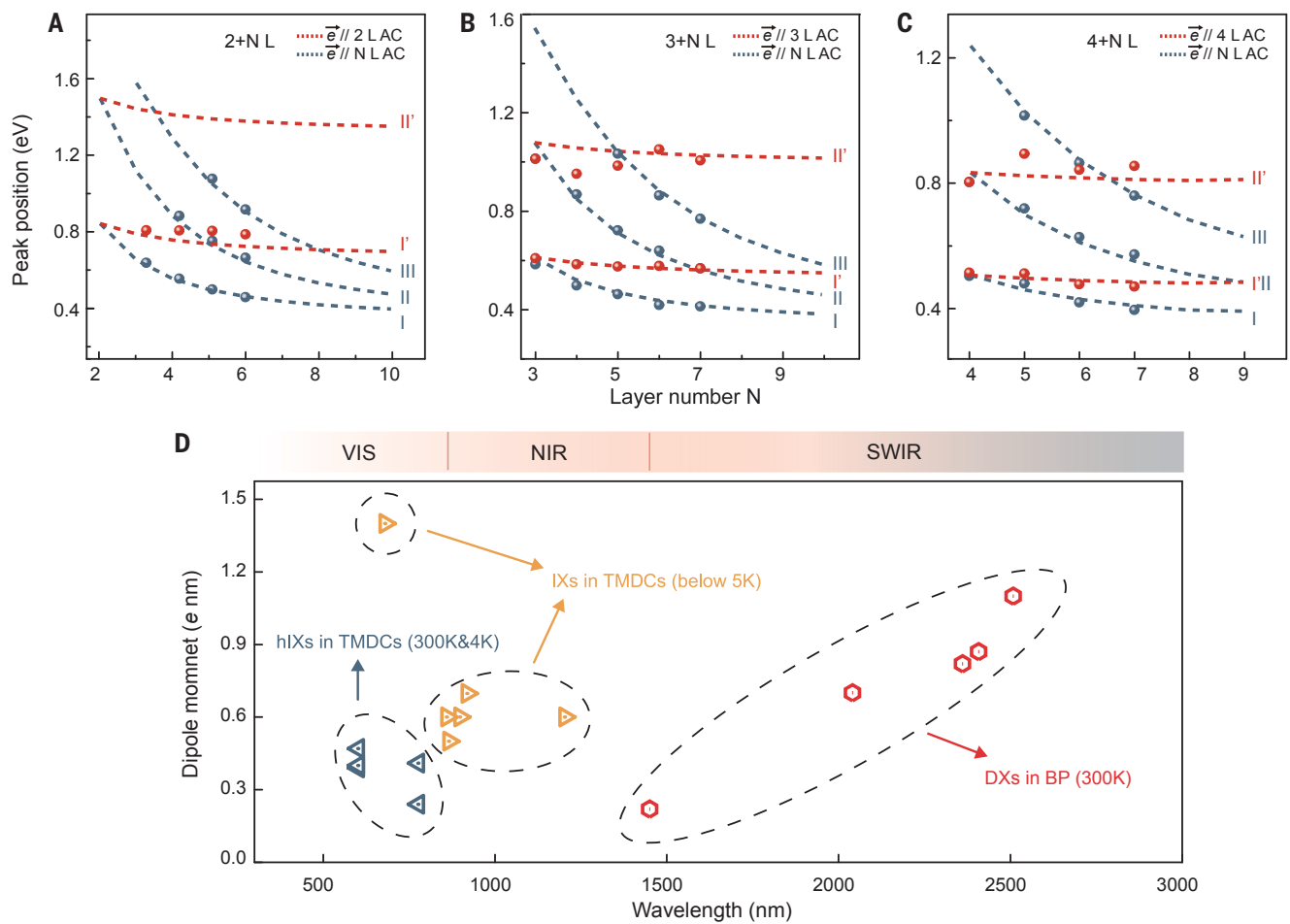

**Fig. 4. Evolution of the dipolar exciton energy with thickness. (A to C)** Energy of dipolar excitons versus layer number N in 90°-twisted BP homostructures composed of 2+N L, 3+N L, and 4+N L BP, respectively. Dots are experimental data. Red and blue dashed curves are fitting curves. **(D)** Comparison of dipolar excitons (DXs), including interlayer excitons (IXs) and hybrid interlayer-intralayer excitons (hIXs), obtained experimentally in TMDC hetero- and homobilayers with those in 90°-twisted BP homostructures (3, 4, 6, 26, 28–32). Theoretically, the dipole moments and working wavelength regime in 90°-twisted BP homostructures can be further extended by choosing different thickness combinations. NIR, near-infrared; SWIR, short-wave infrared; VIS, visible.

## Discussion

We compared the properties of dipolar excitons in BP homostructures with those observed in TMDC hetero- and homostructures (3, 4, 6, 26, 28–32) (Fig. 4D). It is evident that the operational wavelength range of dipolar excitons has been extended into the infrared regime, which will facilitate innovative infrared device applications, such as multidimensional photodetectors (33, 34), light emitters (35–37), gas sensors (38), and emerging microscale spectrometry technologies (39, 40), all benefiting from high electrical tunability as a result of giant Stark effects. Moreover, the dipole moment is highly tunable by adjusting the BP thickness or selecting different interband transitions (e.g., I and II labeled in Fig. 1B), along with polarization-induced switching of dipole orientation. This tunability holds great promise for manipulating dipole-dipole interactions, which is of fundamental importance for nonlinear dipolar polaritons, dipolar complexes (trion, biexciton, etc.), and correlated states (condensation, superfluid, dipolar crystal, etc.) in the field of excitonic physics (41). Notably, the in-plane anisotropy provides these newly found dipolar excitons with much richer physics, such as anisotropic dipolar exciton diffusion, aniso-tropic dipolar polaritons, and so on. Most importantly, our discovery provides a platform for future exploration of tunable correlated states of dipolar excitons in rectangular moiré lattices, which are distinguished from triangular lattices in TMDC moiré systems.

## ACKNOWLEDGMENTS

Funding: H.Y. is grateful for financial support from the National Key Research and Development Program of China (grant nos. 2022YFA1404700 and 2021YFA1400100), the National Natural Science Foundation of China (grant no. 12074085), and the Natural Science Foundation of Shanghai (grant nos. 23XD1400200 and 23JC1401100). K.Y. is grateful for financial support from the National Natural Science Foundation of China (grant no. 12104307). H.W. is grateful for financial support from the National Natural Science Foundation of China (grant nos. 12174062 and 12241402) and from the Innovation Program for Quantum Science and Technology. Part of the experimental work was carried out in the Fudan Nanofabrication Lab. Author contributions: H.Y. and S.H. initiated the project and conceived the experiments. S.H., B.Y., and Yix.M. prepared the samples and performed the measurements with assistance from C.P., Q.X., L.M., J.Z., and Yan.M. S.H., B.Y., and C.P. performed nanofabrication of the devices with assistance from Yix.M. and Y.L. S.H. performed data analysis with assistance from J.M. Y.Z., Yao.M., K.Y., and H.W. provided the DFT calculations. H.Y. and S.H. cowrote the paper, and all authors commented on the paper. H.Y. supervised the whole project. Competing interests: The authors declare no competing interests. Data and materials availability: All data are available in the manuscript or the supplementary materials. License information: Copyright © 2024 the authors, some rights reserved; exclusive licensee American Association for the Advancement of Science. No claim to original US government works. https://www.science.org/about/science-licenses-journal-article-reuse

## SUPPLEMENTARY MATERIALS